RESEARCH ARTICLE

# A novel use of time separation technique to improve flat detector CT perfusion imaging in stroke patients

Vojtěch Kulvait[1,2] | Philip Hoelter[3] | Robert Frysch[1] | Hana Haseljić[1] | Arnd Doerfler[3] | Georg Rose[1]

[1]Institute for Medical Engineering and Research Campus STIMULATE, University of Magdeburg, Magdeburg, Germany

[2]Institute of Materials Physics, Helmholtz-Zentrum Hereon, Geesthacht, Germany

[3]Department of Neuroradiology, University Hospital Erlangen, Friedrich-Alexander-Universität (FAU) Erlangen-Nürnberg, Erlangen, Germany

**Correspondence**
Vojtěch Kulvait, Institute for Medical Engineering and Research Campus STIMULATE, University of Magdeburg, Magdeburg, Germany.
Email: vojtech.kulvait@hereon.de

**Funding information**
Bundesministerium für Bildung und Forschung (BMBF, Germany) within the Research Campus STIMULATE, Grant/Award Number: 13GW0473A

**Abstract**

**Background:** CT perfusion imaging (CTP) is used in the diagnostic workup of acute ischemic stroke (AIS). CTP may be performed within the angio suite using flat detector CT (FDCT) to help reduce patient management time.

**Purpose:** In order to significantly improve FDCT perfusion (FDCTP) imaging, data-processing algorithms need to be able to compensate for the higher levels of noise, slow rotation speed, and a lower frame rate of current FDCT devices.

**Methods:** We performed a realistic simulation of FDCTP acquisition based on CTP data from seven subjects. We used the time separation technique (TST) as a model-based approach for FDCTP data processing. We propose a novel dimension reduction in which we approximate the time attenuation curves by a linear combination of trigonometric functions. Our goal was to show that the TST can be used even without prior assumptions on the shape of the attenuation profiles.

**Results:** We first demonstrated that a trigonometric basis is suitable for dimension reduction of perfusion data. Using simulated FDCTP data, we have shown that a trigonometric basis in the TST provided better results than the classical straightforward processing even with additional noise. Average correlation coefficients of perfusion maps were improved for cerebral blood flow (CBF), cerebral blood volume, mean transit time (MTT) maps. In a moderate noise scenario, the average Pearson's coefficient for the CBF map was improved using the TST from 0.76 to 0.81. For the MTT map, it was improved from 0.37 to 0.45. Furthermore, we achieved a total processing time from the reconstruction of FDCTP data to the generation of perfusion maps of under 5 min.

**Conclusions:** In our study cohort, perfusion maps created from FDCTP data using the TST with a trigonometric basis showed equivalent perfusion deficits to classic CT perfusion maps. It follows, that this novel FDCTP technique has potential to provide fast and accurate FDCTP imaging for AIS patients.

**KEYWORDS**
Brain Stroke, Computed Tomography, CT Perfusion, Model-based Approaches, Perfusion Imaging, Reconstruction Algorithms, Stroke Imaging, Visualization Algorithms

## 1 | INTRODUCTION

CT perfusion imaging (CTP) is a technique used to visualize and quantify abnormal cerebral hemodynamics in acute ischemic stroke (AIS). Numerous recent studies have successfully used perfusion imaging to select stroke patients with large vessel occlusion (LVO), 6 h or more after the onset of the first symptoms, who may still benefit from thrombolysis or endovascular therapy (EVT).[1–4] Based on this evidence, the current guidelines of the American Stroke Association recommend performing CTP in the late time window.[5]

C-arm X-ray units are commonly used in EVT operations for angiography and two-dimensional (2D) imaging. It is actively being researched whether these devices could completely replace CT in the management of patients with AIS.[6–8] Imaging solely with a C-arm flat detector CT (FDCT) in the angio suite could save up to 30 min by eliminating the need to transport the patient to the conventional CT scan.[9,10] The feasibility of FDCT perfusion (FDCTP) imaging in AIS patients has also been investigated.[8,11,12] When compared with CT perfusion, FDCTP suffers from higher noise levels, lower frame rate of the detector, and slower rotation speed of the C-arm. To overcome these drawbacks, there is an ongoing effort to improve FDCTP processing algorithms.[13–19] However, studies on real-world perfusion data are still scarce. In this paper, we specifically investigate the time separation technique (TST)[18] when applied to patient data without prior assumptions about perfusion profiles.

In perfusion imaging, the patient is injected contrast material and the contrast agent dynamics are evaluated over a period of circa 1 min by continual CT scan. These dynamics are described by time attenuation curves (TACs). From TACs, perfusion maps are computed by means of a modified deconvolution algorithm.[20] The TST is a model-based approach that specifically addresses the low number of rotations and the corresponding sparse sampling during FDCTP scanning. The idea is to fit several, typically five to seven, previously known attenuation profiles to the projection data. If these profiles are mutually orthogonal functions, then due to the linearity of the CT operator, it is possible to reconstruct the individually fitted coefficients independently and obtain the TACs as a superposition of the reconstructed values.

One approach to find a basis of functions suitable for the TST is to take attenuation profiles from a large CTP dataset and perform singular value decomposition (SVD), where the first few TAC profiles with the largest singular values are taken as the basis. Since the elements of the singular decomposition are mutually orthogonal, they can be used for dimension reduction within the TST.[18] However, this approach has its limitations. First, the results will always depend on the original CTP dataset, and using a different one will change the results. Second, there is the question of transferability between different centers and protocols. For example, the peak of the arterial input function (AIF) may be shifted in each patient, and the SVD-derived basis might not correctly capture the TACs in situations with a too early or late arrival of the contrast agent. Therefore, in this work, we studied how well the method would perform if we did not use any a priori information about the shape of the TACs, but instead used a general basis of temporal functions.

It would be natural to use gamma variate functions to construct a general basis of the time attenuation profiles. However, for the TST, it is necessary that the basis elements are mutually orthogonal. Unfortunately, there is no natural orthogonal basis for gamma variate functions because the sum of two gamma variate functions is no longer a gamma variate. This is due to their different widths and different positions of the maxima. Therefore, we had to choose a more general orthogonal system. We specifically chose to use the first terms in the trigonometric polynomial as basis functions. These functions are mutually orthogonal and therefore suitable for the TST. Moreover, they appear in the Fourier decomposition and are therefore commonly used in engineering to decompose various signals and time profiles. Hence, we assume that they can also serve as a general basis for decomposing TACs. Our hypothesis was that (i) the basis of trigonometric functions is suitable for the dimension reduction of CTP data and that (ii) the TST with this basis can improve the quality of FDCPT maps.

## 2 | MATERIALS AND METHODS

The processing of perfusion data consists of three major steps:

1. The determination of TACs for each voxel of the scanned volume. In the straightforward approach, used, for example, for CTP data processing, the TACs are modeled as an interpolation of the individual time-resolved volumes. In model-based approaches, such as TST, which are mainly used to process FDCTP data, TACs are approximated by a mathematical model.
2. Localization of the AIF, which is a single TAC that represents the dynamics of the contrast agent at the arterial input site. The selection of the AIF is a critical step because the value of perfusion parameters depends on it.
3. Calculation of perfusion parameters and construction of perfusion maps. For this we used a modified deconvolution model.[20]

Precise algorithms to process perfusion data vary from vendor to vendor.[21–23] Therefore, reproducibility of results is an important issue.[24,25] For the sake of accuracy and to remove ambiguity, we describe extensively the perfusion imaging algorithms, which we use in this paper. We developed and implemented the entire CTP and FDCTP data-processing pipelines, which are able to process a variety of perfusion datasets and were extensively tested using data from various devices.[26,27] Major parts of these pipelines are available as open-source software, in particular algorithms for algebraic reconstruction (https://github.com/kulvait/kct_cbct/)[28] and algorithms for deconvolution processing of

**TABLE 1** Characteristics of the patient cohort. NIHSS refers to the National Institutes of Health Stroke Scale, which is a 42-point scale of the severity of the stroke based on physical examination[30] Note: The modified Rankin scale (mRS).[31] was used to evaluate the degree of disability of the patients at an initial check-up and after 90 days. The scale range goes from, 0 meaning no disability, to 6 meaning death of the patient. The column's location and side describe the location and side of the large vessel occlusion (LVO). ICA stands for internal carotid artery and M1 and M2 stand for respective segments of the middle cerebral artery

| PtsID | Age | LVO | NIHSS | mRS initial | mRS day 90 | Onset to CT (min) | Location | Side | AIF position |
|---|---|---|---|---|---|---|---|---|---|
| 1A | 81 | ✓ | 17 | 3 | 3 | 144 | M1 | R | ICA L |
| 2A | 73 | ✓ | 7 | 2 | 6 | 205 | M2 | L | ICA R |
| 3A | 87 | ✓ | 7 | 1 | 1 | 117 | M2 | L | ICA R |
| 4A | 85 | ✓ | 15 | 0 | 5 | 90 | ICA | R | ICA L |
| 1B | 79 | ✗ | - | - | - | - | - | - | ICA L |
| 2B | 35 | ✗ | - | - | - | - | - | - | ICA L |
| 3B | 87 | ✗ | - | - | - | - | - | - | ICA L |

perfusion data (https://github.com/kulvait/kpct_perfviz/).[29]

## 2.1 | Patient cohort and scanning protocol

Seven randomly selected subjects who underwent a CTP scan for AIS due to suspected LVO between December 2019 and July 2020 were included in this study. All subjects had symptoms of AIS, four of them were confirmed to have LVO and the remaining three without LVO were used as controls. Table 1 summarizes the basic characteristics of the patient cohort, including age, ischemia location, and stroke severity scoring. Informed consent was obtained from patients or their legal representatives in accordance with local laws and regulations. This study was performed in accordance with the ethical standards laid down in the 1964 Declaration of Helsinki and its later amendments.

In the course of perfusion imaging, 30 mL of a contrast agent (Imeron®400, Bracco Imaging, Konstanz, Germany) were injected through an 18-gauge cubital-vein-cannula at a rate of 5 mL s$^{-1}$, followed by 50 mL of saline flush. CTP was performed in the caudocranial direction using a 128-section scanner (Somatom Definition AS+; Siemens Healthcare GmbH, Forchheim, Germany) with a delay of 2 s from the start of contrast injection. Coverage in the z-axis was 93 mm, centered in the basal ganglia (80 kV, 80 mAs). During the CTP scan, 35 volumes were created, the acquisition of one volume took 1.5 s, and the CTP scan took a total of $T_{CT} = 67.98$ s. The dimensions of the volumes were $(n_x, n_y, n_z) = (512, 512, 31)$ and voxel sizes $(\Delta x, \Delta y, \Delta z) = (0.39\text{ mm}, 0.39\text{ mm}, 3.0\text{ mm})$. The short delay between the administration of the contrast agent and the start of acquisition ensures that the first volumes can be considered as a mask without a contrast agent. The onset of the contrast agent in the arteries occurs typically more than 9 s after the start of CT acquisition.

## 2.2 | CTP data and location of the AIF

The CTP scan data in DICOM format contains 35 time-resolved volumes including information on the time of the creation of each slice. To approximate the TAC for each voxel, we interpolate the corresponding attenuation values using Akima splines.[32] Spline interpolation was performed using Intel® Math Kernel Library 2020 for Linux. TACs are shifted to start from zero. This is a natural approach equivalent to subtracting the mask from all time-resolved volumes.

In all subjects (see Table 1), the AIF was located in the internal carotid artery (ICA) near the skull base. In LVO patients, the ICA opposite to the occlusion side was used. In control subjects, the AIF was located in the left ICA. Based on these criteria, the first two authors determined the position of the AIF by common consensus. The second author, a trained radiologist specializing in neuroradiology, was responsible for the anatomical positioning of the AIF, and the first author, from the perspective of a medical data scientist, was responsible for the correct shape of the attenuation curve and the precise determination of the voxel corresponding to the maximum flow rate of the contrast agent.

## 2.3 | Simulated FDCTP data with additional noise

For the purpose of this work, we simulate the following C-arm FDCT perfusion protocol based on current experimental FDCTP protocols for brain perfusion.[15,16,18] During FDCTP acquisition, the C-arm of the FDCT scanner performs five consecutive pairs of back and forth rotations to capture the dynamics of the contrast agent and to acquire projections (see figure 1). Each of $R = 10$ unidirectional rotations is called a sweep. One sweep takes 4.1 s, and the time gap between two consecutive sweeps is $T_p = 2.3$ s (see figure 2). The projection data for each sweep contains $Q = 248$ projections with the flat detector size $m_1 \times m_2 = 616 \times 480$ and the pixel

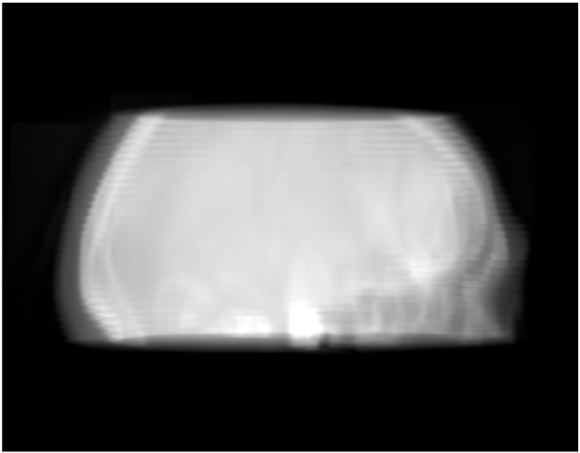

**FIGURE 1** FDCTP reprojected data of patient 1A, a single view of the 212-th angle of the first sweep using the flat panel detector with dimensions 616 × 480. The original discretization of the CTP reconstructions causes step-like patterns in the reprojected data. The FDCT field of view is larger compared with the CT acquisition

size $0.616 \times 0.616$ mm$^2$. For the time synchronization with CTP acquisition, we assume that there is a delay of 3.2 s between the start of contrast injection and the start of FDCTP acquisition. The end of FDCTP acquisition is 2 s before the end of the CTP acquisition. The FDCTP scan covers the entire dynamics of the contrast agent, and the first sweep can be considered as a mask (see figure 3). To realistically simulate FDCTP acquisition and obtain the projection data corresponding to the 10 rotations of the C-arm CT, we use CTP datasets of the studied patient cohort. We interpolate 35 time-resolved attenuation volumes according to the temporal dynamic model described in Section 2.2 In this way, we can compute any volume that corresponds to the attenuation at a given time instant $t$ within the CTP acquisition interval $I = [0, T_{CT}]$. Simulated FDCTP scan interval is $\mathcal{I} = [T_1, T_2]$, where $0 < T_1 < T_2 < T_{CT}$. Therefore the time of given projection $t_q^r$, where $q \in \{0 \ldots 247\}$ is a projection index and $r \in \{0 \ldots 9\}$ is a sweep index can be computed using the formula

$$t_q^r = T_1 + (T_p + 247 * T_f) * r + q * T_f, \quad (1)$$

where $T_1 = 1.2$ s is the start of FDCTP acquisition, $T_p = 2.3$ s is the pause size between rotations and $T_f = 16.7$ ms is a delay between two consecutive views in one sweep. Total FDCTP scan time is $T = T_2 - T_1 = 62.1$ s. We compute all 2480 volumes $V_q^r$ corresponding to the times $t_q^r$ of every view. Then we use a CT projector[1] to reproject individual volumes $V_q^r$ and obtain 2480 FDCTP projections on the 616 × 480 detector (see figure 1). Using this reprojection technique, we obtain the complete FDCTP projection data that are qualitatively similar and have the same dimensions as the real FDCTP datasets.

Typically, FDCTP data contain more noise than CTP data. To increase the realism of the reprojected FDCTP data, we added two different noise levels to the FDCTP projections. In the moderate noise scenario,[16] Poisson noise was added to the projection data corresponding to the acquisition of a blank image with 6×10$^5$ photons per mm$^2$. In the high noise scenario,[15] added Poisson noise corresponds to 2.1×10$^5$ photons per mm$^2$.

A step-like pattern appears in the simulated FDCTP data (see figure 1). This problem is caused by the original discretization of CTP volumes, where the distance between z-slices is $\Delta z = 3.0$ mm. To compensate for this and to avoid introducing additional artifacts, we use the same volume discretization in the reconstruction of the FDCTP data as used in the discretization of the original CTP data.

## 2.4 | Estimating TACs based on FDCTP data

Processing FDCTP data involves CT reconstructions of the reprojected data. To avoid limited view artifacts in the straightforward processing of FDCTP data, it is natural to reconstruct 10 individual sweeps separately (see figure 2). In straightforward processing, we perform the interpolation based on these 10 volumes analogous to the processing in Section 2.2, but instead of interpolating 35 time-resolved volumes, we interpolate just 10. Straightforward processing therefore involves 10 individual CT reconstructions.

In the TST, we first fit a few temporal profiles, for example, trigonometric functions, to the FDCTP projection data and then reconstruct coefficients corresponding to these profiles. The number of reconstructions is the same as the number of temporal profiles (basis functions), typically five. In perfusion imaging, algebraic reconstruction generates less noisy results compared to analytical reconstruction.[18] Since the fitting of temporal profiles in the TST amplifies the noise in the data due to fitting errors, classical algebraic reconstruction techniques often diverge, especially in higher terms of the fitted basis. Therefore, for CT reconstruction we use the implementation[28] of the CGLS algorithm. CGLS is a method for solving large systems of linear equations derived from the conjugate gradients algorithm applied to normal equations. This method provides enhanced stability and convergence properties even for extremely noisy data,[34] so that divergence of solutions is not encountered even when reconstructing coefficients with high fitting error. For the full description of the CGLS algorithm and a broader discussion of its use for tomographic reconstruction, see the paper by Kulvait and Rose.[34]

---

[1] For CT projections, we use the implementation[28] of the TT projector, which is the most accurate cone beam CT projector from the class of the so-called footprint projectors.[33]

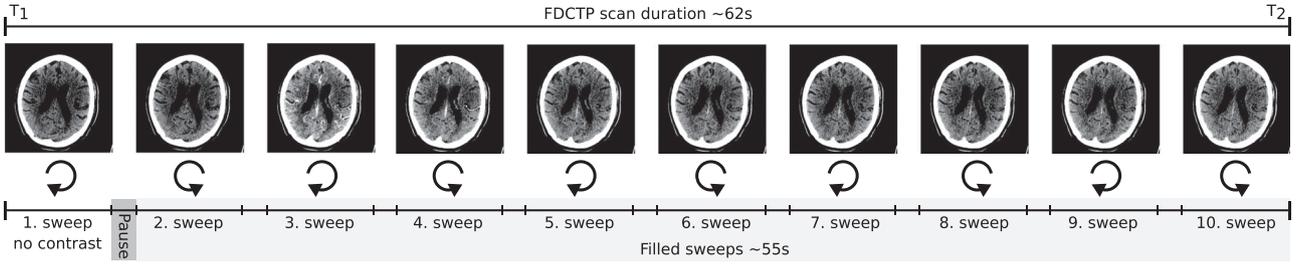

**FIGURE 2** Scheme of the simulated brain FDCTP protocol

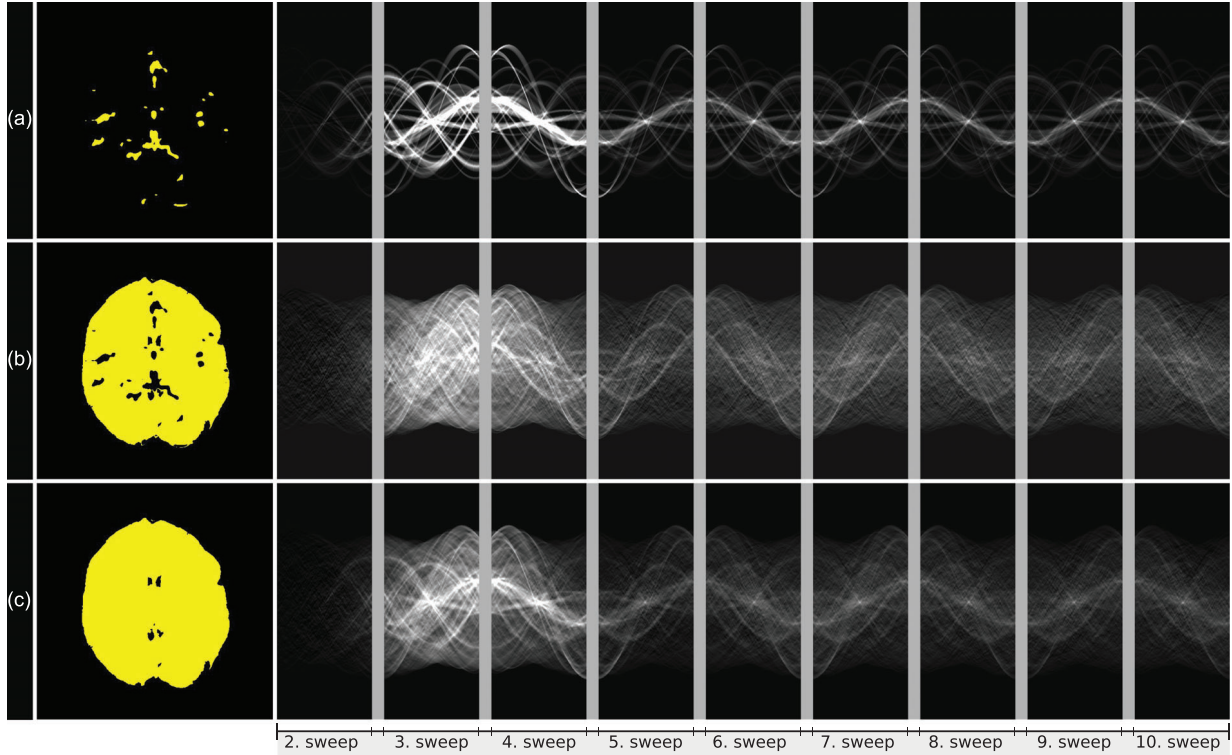

**FIGURE 3** Simulated sinogram for FDCTP reprojected data describing the dynamics of the contrast agent in the central slice of control subject 2B over nine contrast-enhanced sweeps, where the first sweep is used as a subtraction mask. Additional tissue masks were used to separate dynamics in large vessels from general soft tissue dynamics. A: vessel mask, B: soft tissues without large vessels, C: soft tissues including large vessels

## 2.5 | Mathematical derivation of TST

When reconstructing the FDCTP projection data, we deal with a time-dependent problem where both the reconstructed volumes and the projection data are functions of time. Volume is discretized by $N = n_x \times n_y \times n_z = 512 \times 512 \times 19$ voxels, the detector has $m = m_1 \times m_2 = 616 \times 480$ pixels and there is total $M = Q \times m = 248 \times 616 \times 480$ pixel values read during the single sweep. Let the time-dependent vector $\mathbf{x}(t) : \mathcal{I} \to \mathbb{R}^N$ represent the scanned volume at time $t$, which is to be reconstructed. We assume that for each configuration $q$ there is a projection operator $\mathbf{A}_q \in \mathbb{R}^{m \times N}$ that projects the volume $\mathbf{x}(t) \in \mathbb{R}^N$ onto the detector extinction vector $\mathbf{p}_q(t) \in \mathbb{R}^m$ such that

$$\mathbf{A}_q \mathbf{x}(t) = \mathbf{p}_q(t), \quad t \in \mathcal{I}. \quad (2)$$

We can assemble all operators $\mathbf{A}_q$ and all projection vectors $\mathbf{p}_q(t)$ to get the global operator $\mathbf{A} \in \mathbb{R}^{M \times N}$ and the global projection vector $\mathbf{p}(t) : I \to \mathbb{R}^M$. It holds that

$$\mathbf{A} = \begin{pmatrix} \mathbf{A}_1 \\ \vdots \\ \mathbf{A}_Q \end{pmatrix}, \quad \mathbf{p}(t) = \begin{pmatrix} \mathbf{p}_1(t) \\ \vdots \\ \mathbf{p}_Q(t) \end{pmatrix}, \quad \mathbf{A}\mathbf{x}(t) = \mathbf{p}(t), \quad t \in \mathcal{I}.$$

$$(3)$$

Equation (3) describes a time-resolved CT problem. For a given instant of time, we know only part of the values of the vector **p**(t) corresponding to the geometric configuration of the CT at that instant. We do not know the values for the other configurations and therefore cannot use (3) for a direct reconstruction to obtain **x**(t). In the straightforward approach, we violate the principal assumption of the CT reconstruction that the scanned object is static and use the projection data acquired during the interval $[t_0^r, t_{247}^r]$ of $r$-th sweep to reconstruct a volume $\mathbf{x}^r \approx \mathbf{x}(t^r)$, where $t^r = (t_0^r + t_{247}^r)/2$. Then we interpolate the resulting 10 volumes analogously to the CTP data processing (see Section 2.2).

In order not to violate the principal assumption of the CT reconstruction and still use the time-resolved CT model (3), we have to use a model-based approach.[13,14,18] In the TST, we assume that both projection data and reconstructed volumes can be modeled as a linear combination of a set of $K$ temporal functions. Let us suppose that

$$\mathcal{B} = \{\Psi_1, \ldots, \Psi_K\}$$

is the set of temporal profiles $\Psi_i = \Psi_i(t)$, $t \in \mathcal{I}$, $i \in \{1, \ldots, K\}$. We assume that the functions $\Psi_i$ are from the vector space $V$ with a scalar product $\langle \cdot, \cdot \rangle$, and that they are mutually orthogonal, so $\langle \Psi_i, \Psi_j \rangle = 0$ for $i \neq j$. Using these assumptions, we rewrite both the time-dependent volume

$$\mathbf{x}(t)(x_1, x_2, x_3) = \sum_{k=1}^{K} \mathbf{w}^k(x_1, x_2, x_3) \Psi_k(t), \quad (4)$$

and the time-dependent projection data

$$\mathbf{p}_q(t)(m_1, m_2) = \sum_{k=1}^{K} \mathbf{c}_q^k(m_1, m_2) \Psi_k(t), \quad (5)$$

as a linear combination of basis functions. Using Equations (4) and (5), the time-resolved CT problem (3) takes the form

$$\mathbf{A} \sum_{k=1}^{K} \mathbf{w}^k \Psi_k(t) = \sum_{l=1}^{K} \mathbf{c}^l \Psi_l(t), \quad t \in \mathcal{I}. \quad (6)$$

If we take an element-wise scalar product of both sides of Equation (6) with the functions $\Psi_i, i \in \{1 \ldots K\}$, we obtain

$$\mathbf{A} \sum_{k=1}^{N} \mathbf{w}^k \langle \Psi_k, \Psi_i \rangle = \sum_{l=1}^{N} \mathbf{c}^l \langle \Psi_l, \Psi_i \rangle, \quad (7)$$

which results, due to the orthogonality of the basis, in the $K$ independent standard static reconstruction problems of the form

$$\mathbf{A} \mathbf{w}^i = \mathbf{c}^i, \quad i \in \{1 \ldots K\}. \quad (8)$$

Thus, under the TST assumptions, we can recover the full temporal dynamics by solving $K$ static reconstruction problems.

In order to use (8) to compute the coefficients $\mathbf{w}^i$ and recover the volume dynamics (4), we need to estimate the coefficients $\mathbf{c}^i$, $i \in \{1 \ldots K\}$ describing the dynamics of the projection data (5). To do this, we use standard regression analysis using given basis functions. For each view, $q$ we have the number of measurements equal to the number of sweeps. Specifically, we have that $\mathbf{p}_q(t_q^r) = \mathbf{p}_q^r$ for $r \in \{0 \ldots 9\}$ and we can fit these values to Equation (5) to obtain the vectors $\mathbf{c}^l, l \in \{1 \ldots K\}$ as the least-squares solution. To obtain TACs from attenuation dynamics, we subtract $\mathbf{p}_q(0)$ from (5) so that TACs begin from zero.

In this work, we have chosen a basis consisting of mutually orthogonal trigonometric functions

$$\Psi_0 = 1, \quad \Psi_1 = \sin\left(\frac{2\pi t}{T}\right), \quad \Psi_2 = \cos\left(\frac{2\pi t}{T}\right),$$

$$\Psi_3 = \sin\left(\frac{4\pi t}{T}\right), \quad \Psi_4 = \cos\left(\frac{4\pi t}{T}\right),$$

$$\Psi_5 = \sin\left(\frac{6\pi t}{T}\right), \quad \Psi_6 = \cos\left(\frac{6\pi t}{T}\right). \quad (9)$$

We use either the basis $\mathcal{T}_4 = \{\Psi_0, \Psi_1, \Psi_2, \Psi_3, \Psi_4\}$ or $\mathcal{T}_6 = \{\Psi_0, \Psi_1, \Psi_2, \Psi_3, \Psi_4, \Psi_5, \Psi_6\}$. Since $\Psi_0$ is constant, we refer to $\mathcal{T}_4$ and $\mathcal{T}_6$ as the bases of four and six trigonometric coefficients. Therefore, we represent the data by a low-order trigonometric polynomial with a period $T$, which corresponds to the duration of the FDCTP scan.

## 2.6 | Generating perfusion maps by deconvolution algorithm

The perfusion data obtained by the techniques described above consist of attenuation profiles $x_v(t)$, $t \in \mathcal{I}$, where $v$ is the voxel index. Let us denote TACs by $y_v$, where $y_v(t) = x_v(t) - x_v(T_1)$, $t \in \mathcal{I}$. TACs are therefore shifted attenuation profiles that begin at zero. The perfusion parameters, namely cerebral blood flow CBF, cerebral blood volume CBV, mean transit time MTT and time to peak TTP, represent the aggregate information about the perfusion in a given voxel calculated from the perfusion data. Mathematically, these parameters can be formalized using functionals that assign a single real number representing the value of a given perfusion parameter to a function $y_v(t)$. Calculating the perfusion parameters for the entire imaged volume yields perfusion maps, the diagnostic output of perfusion imaging.

Here we briefly outline the method of computing perfusion parameters based on the indicator dilution theory.[20] To use this theory, we need to discretize TAC $y(t)$ by the vector $\mathbf{y} \in \mathbb{R}^G$, which represents its values in $G$ equidistant points of the interval $[T_1, T_2]$ such that the $i$th component of the vector $\mathbf{y}$ is

$$\mathbf{y}^i = y(T_1 + \Delta(i-1)), \quad \Delta = \frac{T_2 - T_1}{G - 1}, \quad i \in \{1 \cdots G\}. \tag{10}$$

In this study, the interval granularity was set to $G = 100$. It is a compromise between the density of the sampling that is 0.62 s and the speed of deconvolution processing. Calculations of all perfusion parameters except TTP depend on the AIF described by $\mathbf{y}_{AIF} \in \mathbb{R}^G$. Therefore, the selection of AIF is essential for the quality of perfusion imaging. Let the discretized TAC for a given voxel be given by the function $\mathbf{y}_v \in \mathbb{R}^G$. It is assumed that there exists a convolution kernel $\mathbf{k}_v \in \mathbb{R}^G$ such that $\mathbf{y}_v = \mathbf{y}_{AIF} * \mathbf{k}_v$, where $*$ is a discrete convolution operator for which the following equation holds:

$$\begin{pmatrix} \mathbf{y}_v^1 \\ \mathbf{y}_v^2 \\ \vdots \\ \mathbf{y}_v^G \end{pmatrix} = \mathbf{Y}_{AIF} \begin{pmatrix} \Delta \mathbf{k}_v^1 \\ \Delta \mathbf{k}_v^2 \\ \vdots \\ \Delta \mathbf{k}_v^G \end{pmatrix}, \quad \text{where}$$

$$\mathbf{Y}_{AIF} = \begin{pmatrix} \mathbf{y}_{AIF}^1 & 0 & 0 & \cdots & 0 \\ \mathbf{y}_{AIF}^2 & \mathbf{y}_{AIF}^1 & 0 & \cdots & 0 \\ \vdots & \vdots & \vdots & \ddots & \vdots \\ \mathbf{y}_{AIF}^G & \mathbf{y}_{AIF}^{G-1} & \mathbf{y}_{AIF}^{G-2} & \cdots & \mathbf{y}_{AIF}^1 \end{pmatrix}. \tag{11}$$

The Toeplitz matrix $\mathbf{Y}_{AIF}$ is a discrete version of the convolution operator. Theoretically, by inverting the matrix $\mathbf{Y}_{AIF}$, we can obtain the convolution kernel using the equation $\Delta \mathbf{k}_v = \mathbf{Y}_{AIF}^{-1} \mathbf{y}_v$. The matrix $\mathbf{Y}_{AIF}$ is however compromised by noise.[20] Therefore, the computation of the $\mathbf{Y}_{AIF}^{-1}$ is an ill-conditioned problem that needs to be stabilized. We need to invert parts of $\mathbf{Y}_{AIF}$ corresponding to large singular values and to diminish the influence of extremely small singular values by constructing the matrix $\mathbf{Y}_{AIF}^{INV} \approx \mathbf{Y}_{AIF}^{-1}$. The technique with modified matrix $\mathbf{Y}_{AIF}$ to a block-circulant form is often used for processing MRI perfusion data.[35] In this paper, we use Tikhonov stabilization of the $\mathbf{Y}_{AIF}$ matrix with parameter $\lambda_{rel} = 0.2$, as proposed in Fieselmann et al.[20] First, we construct the matrix $\mathbf{Y}_{AIF}^{INV}$ based on AIF. Then for each TAC $\mathbf{y}_v \in \mathbb{R}^G$ we compute the convolution kernel $\Delta \mathbf{k}_v = \mathbf{Y}_{AIF}^{INV} \mathbf{y}_v$ and finally the perfusion parameters for a given TAC are computed as follows.

TTP is computed as a time from the start of the acquisition when the maximum TAC value is reached.

CBF is, according to indicator dilution theory, expressed as the first element of the convolution kernel $\mathbf{k}_v^1$. This corresponds, based on Equation (11), to how close the TAC resembles the shape of the AIF without any delay. In Fieselmann et al.[20, (p. 9)], the authors noticed that taking the first element of the vector $\mathbf{k}_v$ leads to instabilities and suggest using its maximum value instead. This choice has apparently been implemented in some perfusion processing software. By mathematical intuition, we now seek the value of $\mathbf{k}_v$ at time when the shape of the TAC most closely resembles the AIF, irrespective of the delay. We observe that this makes the dynamics of the CBF very similar to that of the CBV, making their ratio, MTT, mostly redundant. Since MTT is supposed to be an indicator of collateral flow, the algorithm in Fieselmann et al.[20, (p. 9)] could undermine the quality of therapeutic decisions. By taking the maximum of $\mathbf{k}_v$ in the CBF calculation, MTT would not accurately capture the flow delay, and the perfusion deficit could be erroneously assessed as not being accompanied by collateral reconstitution. On the other hand, taking $\mathbf{k}_v^1$ could lead to an underestimation and the instability of CBF due to noise and the presence of a natural flow delay between the arterial input and the measured parenchymal location. A compromise approach used in this work is to calculate the maximum of the convolution kernel over a short period at the beginning of the acquisition. We set the parameter $T_{CBF}^{MAX} = 5$ s and compute the CBF as follows:

$$\text{CBF}(\mathbf{y}_v) = \max\{\mathbf{k}_v^i, \quad i \in \{1 \cdots G\}, \quad (i-1)\Delta \leq T_{CBF}^{MAX}\}. \tag{12}$$

Setting $T_{CBF}^{MAX} = T$ in (12) yields CBF calculation according to Fieselmann et al.,[20] while setting $T_{CBF}^{MAX} = 0$ leads to the theoretically optimal value from indicator dilution theory.

CBV is related to the amount of blood that flows through a given voxel. We compute it by integration or, in the discrete case, by summing all values of the vector $\mathbf{k}_v$

$$\text{CBV}(\mathbf{y}_v) = \sum_{i=1}^{G} \mathbf{k}_v^i. \tag{13}$$

MTT refers to the time it takes for a unit of blood with a contrast medium to be transferred from the arterial input to a given point. Elevated values of MTT at normal CBV may refer to collateral flow reconstitution. It is calculated as the ratio of CBV to CBF

$$\text{MTT}(\mathbf{y}_v) = \frac{\text{CBV}(\mathbf{y}_v)}{\text{CBF}(\mathbf{y}_v)}. \tag{14}$$

## 2.7 | Brain segmentation and precise location of AIF

When visualizing perfusion in soft tissues of the brain, it is important to exclude any signal that would impair the readability of the resulting perfusion maps. For segmentation, we first create a soft tissue mask and exclude areas outside the intracranial region. We then remove large vessels from the visualization, as high signals from vessels would impair the contrast of the perfusion data in the parenchyma.

To create a soft tissue mask, we use thresholding based on a range of Hounsfield units [20,100] combined with the boundary fill algorithm with the seed inside the intracranial area. The boundary fill algorithm is an algorithm for filling closed polygons based on specified criteria. We use it to select an intracranial region based on the range of Hounsfield units, so as not to select regions outside the skull or inside the hollow bones that would normally be selected by the thresholding algorithm alone (see figure 3c). In the straightforward method, we segment based on the average of the reconstructed volumes and in the case of TST we use a coefficient corresponding to a constant basis function $\Psi_0 = 1$ to guide the segmentation. This choice of volumes reduces noise in both CTP and FDCTP data.

To exclude large blood vessels from the visualization, we can create a mask based on the one percentile of the highest intracranial CBF signal. The CBF signal is an order of magnitude higher in vessels than in soft tissues. It has been shown recently, that a high CBF signal is a robust predictor that can be used in segmenting the cerebral vasculature.[27] The drawback of this approach is that segmentation would implicitly depend on the position of the AIF from which the CBF signal is calculated. Therefore, we introduce the CBFGV parameter, which is calculated as the CBF from Equation (12), based on the artificial AIF given by the gamma variate function

$$\mathbf{y}_{GV}(t) = \frac{t}{t_{max}} \exp\left(1 - \frac{t}{t_{max}}\right), \quad t \in [0, T], \tag{15}$$

where $t_{max}$ is the peak time.[36] In this study, we set $t_{max} = 9$ s. The high CBFGV values in the perfusion data correspond to the signal of large vessels and therefore we use the upper percentile of this signal for their segmentation (see figure 3a,b).

As CBFGV can be calculated prior to AIF selection, this parameter can also be a guide for exact AIF localization. At the beginning of the Materials and Methods section, we described that we localize the AIF in an artery near the skull base. However, in this region, the signal of the artery may be confounded by a high bone signal. Therefore, CBFGV maps were also presented to the neuroradiologist to guide the decision on the AIF location. The maximum CBFGV in the selected artery was used to determine the exact location of a particular AIF voxel. In the future, a similar approach could be used to semiautomatically localize AIF or also as input to a machine-learning algorithm.

**TABLE 2** Pearson's correlation coefficients comparing ground truth CTP data with CTP data of reduced dimension using $\mathcal{T}_4$ basis (top) and $\mathcal{T}_6$ basis (bottom)

|  | Patient | | | | Control | | |
| --- | --- | --- | --- | --- | --- | --- | --- |
|  | 1A | 2A | 3A | 4A | 1B | 2B | 3B |
| CBF | 0.95 | 0.94 | 0.96 | 0.98 | 0.96 | 0.97 | 0.98 |
| CBV | 0.94 | 0.86 | 0.89 | 0.91 | 0.89 | 0.93 | 0.94 |
| MTT | 0.81 | 0.76 | 0.75 | 0.77 | 0.75 | 0.74 | 0.75 |
| TTP | 0.48 | 0.57 | 0.54 | 0.55 | 0.56 | 0.41 | 0.46 |
| CBF | 0.96 | 0.96 | 0.97 | 0.98 | 0.97 | 0.99 | 0.98 |
| CBV | 0.94 | 0.87 | 0.90 | 0.94 | 0.92 | 0.92 | 0.95 |
| MTT | 0.84 | 0.79 | 0.77 | 0.80 | 0.78 | 0.76 | 0.82 |
| TTP | 0.51 | 0.65 | 0.61 | 0.62 | 0.64 | 0.54 | 0.52 |

## 2.8 | Visualization and additional Gauss blur

For visualization of perfusion maps, we use the color map designed by the Acute Stroke Imaging Standardization Group Japan.[37] To reduce noise in the data and to ensure comparability with the results of the major perfusion packages, we add the Gauss blur with $\sigma = 3.5$ px $= 1.365$ mm to all slices of perfusion maps regardless of the type of map and processing algorithm.

## 2.9 | Correlation of perfusion maps

We use Pearson's correlation coefficient to quantify the differences between pairs of perfusion maps produced by different approaches. Initially, we apply a brain soft tissue mask to both maps being compared, excluding large vessels (see Section 2.7). Then, we correlate the pairs of corresponding values in both volumes. Based on this approach, we create tables with correlations of perfusion parameters CBF, CBV, MTT, and TTP for all study subjects (see Tables 2–4).

## 3 | RESULTS

We consider the CTP data processed by interpolating 35 time-resolved volumes according to Section 2.2 being ground truth data. Figure 5a shows an example slice of the ground truth perfusion maps for Patient 1A.

**TABLE 3** Pearson's correlation coefficients comparing ground truth CTP data with moderate noise FDCTP data processed by the straightforward method (top) and with moderate noise FDCTP data processed by TST with $\mathcal{T}_4$ basis (bottom)

|     | Patient ||||  Control |||
| --- | --- | --- | --- | --- | --- | --- | --- |
|     | 1A | 2A | 3A | 4A | 1B | 2B | 3B |
| CBF | 0.65 | 0.83 | 0.75 | 0.70 | 0.76 | 0.87 | 0.79 |
| CBV | 0.80 | 0.74 | 0.74 | 0.71 | 0.77 | 0.81 | 0.81 |
| MTT | 0.38 | 0.36 | 0.39 | 0.39 | 0.38 | 0.33 | 0.37 |
| TTP | 0.36 | 0.39 | 0.32 | 0.41 | 0.36 | 0.25 | 0.26 |
| CBF | 0.77 | 0.76 | 0.82 | 0.76 | 0.84 | 0.88 | 0.87 |
| CBV | 0.83 | 0.71 | 0.75 | 0.73 | 0.79 | 0.80 | 0.83 |
| MTT | 0.56 | 0.43 | 0.46 | 0.47 | 0.41 | 0.37 | 0.45 |
| TTP | 0.31 | 0.36 | 0.33 | 0.37 | 0.31 | 0.23 | 0.29 |

**TABLE 4** Pearson's correlation coefficients comparing ground truth CTP data with high noise FDCTP data processed by the straightforward method (top) and with high noise FDCTP data processed by TST with $\mathcal{T}_4$ basis (bottom)

|     | Patient ||||  Control |||
| --- | --- | --- | --- | --- | --- | --- | --- |
|     | 1A | 2A | 3A | 4A | 1B | 2B | 3B |
| CBF | 0.56 | 0.76 | 0.64 | 0.67 | 0.63 | 0.83 | 0.70 |
| CBV | 0.75 | 0.70 | 0.66 | 0.67 | 0.71 | 0.76 | 0.75 |
| MTT | 0.28 | 0.28 | 0.29 | 0.33 | 0.30 | 0.28 | 0.28 |
| TTP | 0.29 | 0.30 | 0.24 | 0.32 | 0.28 | 0.20 | 0.19 |
| CBF | 0.72 | 0.72 | 0.76 | 0.74 | 0.80 | 0.85 | 0.82 |
| CBV | 0.80 | 0.67 | 0.71 | 0.71 | 0.75 | 0.78 | 0.80 |
| MTT | 0.45 | 0.35 | 0.36 | 0.39 | 0.34 | 0.33 | 0.35 |
| TTP | 0.27 | 0.29 | 0.27 | 0.31 | 0.25 | 0.19 | 0.22 |

AIF for a single study subject was determined by a single voxel according to Section 2. To ensure comparability of results, the same voxel was chosen when processing the CTP and FDCTP data. In figure 4 top, for patient 1A, the comparison is shown of the AIF reconstructed from the CTP data with the AIF reconstructed from the FDCTP datasets with different levels of the noise (see Section 2.3). In figure 4 bottom, for patient 1A, a comparison is displayed of the AIF reconstructed using CTP data with the reduced dimension and using TST on the FDCTP data with the basis of the four trigonometric functions $\mathcal{T}_4$. The baseline value in Hounsfield units at the origin of the charts in figure 4 is included for convenience. In the actual algorithm for calculating perfusion parameters, only the dynamics of the contrast agent are used and the baseline is subtracted from the data (see Section 2.6).

## 3.1 | Dimension reduction for CTP data

The TST is a dimension reduction method where we first fit given temporal functions to the projection data and then reconstruct time attenuation curves in the imaged volume. Therefore, this method is burdened with two different types of errors. The first error is due to the fact that the actual TACs differ from their approximation by the linear combination of basis functions. The second is the fitting error due to sparsity and the noise in the projection data. Even if the first error is substantial, actual inaccurate approximation by the linear combination of basis functions might still preserve important properties of the perfusion maps. Therefore, we study how these errors affect the quality of the resulting perfusion maps. In particular, we study if the trigonometric functions are suitable for generating plausible perfusion maps when we fit them directly to the CTP data without TST reprojection. It is the prerequisite to be able to the use given basis in the scope of the TST processing of FDCTP data. By means of regression analysis and least-squares fitting, the dimension of the original CTP data was reduced using the bases $\mathcal{T}_4$ and $\mathcal{T}_6$ of four and six trigonometric functions, respectively, which are defined in (9). We obtain the volume coefficients related to the particular basis functions in (4). Table 2 shows the correlation coefficients between the perfusion maps generated using the ground truth data and the reduced dimension data using the $\mathcal{T}_4$ and $\mathcal{T}_6$ bases of the four and six trigonometric functions, respectively. By comparing figure 5a with figure 5b, the difference between the ground truth CTP maps and the maps generated from the data of reduced dimension using the $\mathcal{T}_4$ basis can be seen in the example visualization of Patient 1A.

## 3.2 | Comparison of TST with the straightforward method

Based on Section 2.9, a quantitative analysis of the FDCTP data processed by the straightforward method when interpolating 10 reconstructed volumes from individual sweeps and by the TST with the basis of trigonometric functions was performed. Tables 3 and 4 compare the ground truth CTP data to the FDCTP data processed by the straightforward method (left) and to the same FDCTP data processed by TST with the $\mathcal{T}_4$ basis (right). Table 3 shows comparisons under the moderate noise scenario, while Table 4 contains analogous comparisons for FDCTP data with high noise (see Section 2.3)

## 3.3 | Run times

CT reconstruction problems were solved using the OpenCL implementation of the CGLS method.[34] using

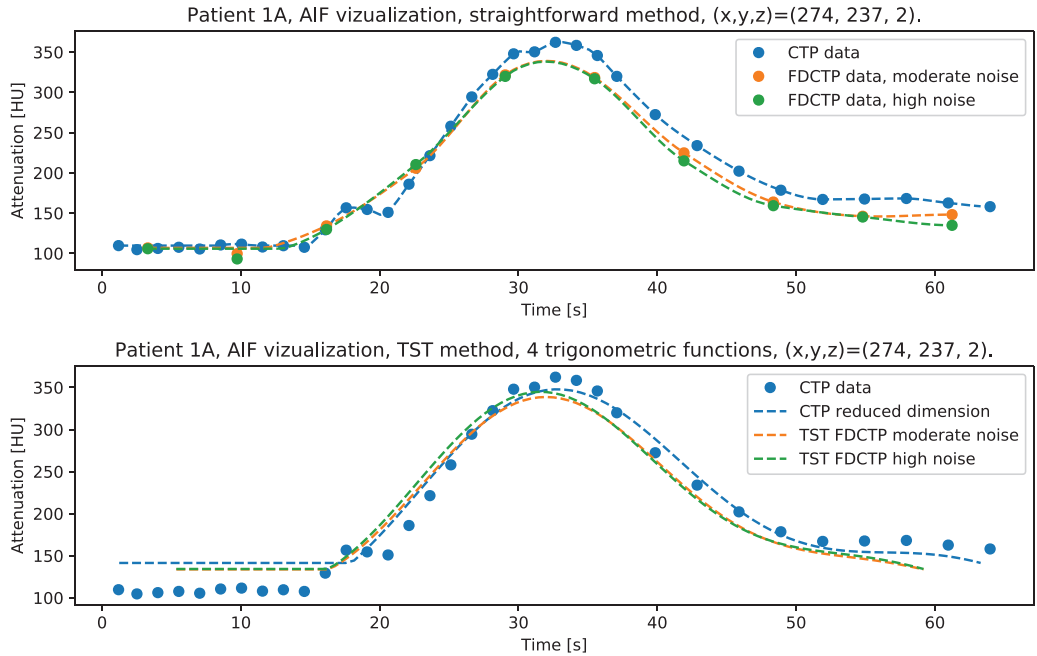

**FIGURE 4** Patient 1A. AIF is derived from spline interpolation of CTP data and straightforward processing of FDCTP data (top). AIF derived from reduced dimension CTP data and from FDCTP data processed using the TST with $\mathcal{T}_4$ basis (bottom)

five RTX2080 Ti GPUs. Other computations were performed on a machine with four Intel® Xeon E7-8890 CPUs. We use SSDs and parallel processing to achieve the highest possible speed. We compare the processing of FDCTP data using TST with the $\mathcal{T}_4$ basis, which contains a total of five basis functions, with the straightforward method and report the average running times. Data preprocessing took 17 s for both the TST and the straightforward method. Fitting the five basis functions to the data within the TST took 71 s. The reconstruction took 128 s for the TST and 275 s for straightforward method. The generation of perfusion maps from the reconstructed data using the deconvolution method took 81 s for the TST and 55 s for the straightforward method. We achieved a total run time of 4 min 57 s for the TST processing and 5 min 47 s for the straightforward method.

## 4 | DISCUSSION

Studying the dimension reduction of the original CTP data in (4) using the basis $\mathcal{T}_4$ of four trigonometric functions, we see that perfusion maps preserve important diagnostic information (see figure 5a,b and Table 2). An interesting finding was that the quality improvement when using the basis $\mathcal{T}_6$, which contains two extra-temporal functions compared with the basis $\mathcal{T}_4$, was negligible both in terms of the visual quality of the resulting maps (see figure 5c), and in terms of quantitative analysis (see Table 2).

The TST is a dimension reduction method where all TACs are reduced to a superposition of $K$ temporal profiles, where $K$ must be smaller than the number of rotations of the FDCT machine, in order to fit the data. As the number of $K$ increases, the number of reconstructions according to Equation (8) also increases, and it is therefore important to balance the number of basis functions so that they are able to capture the behavior of the contrast agent and produce accurate perfusion maps, and on the other hand to keep this number small enough to reduce noise and computational complexity. Similar to the previous paragraph, a pattern of only small improvement was observed when we used a TST with a basis $\mathcal{T}_6$ instead of basis $\mathcal{T}_4$ to process the FDCTP data, data not shown here due to lack of space. It therefore appears that the choice of the trigonometric basis $\mathcal{T}_4$ is optimal, so we focus on presenting results using this basis.

When comparing the AIF profiles of ground truth CTP data with the AIF profiles of FDCTP data calculated by the straightforward method (see figure 4 top), it is apparent that the imprecision of the AIF estimation when using the FDCT is relatively small. The attenuation profiles of the arterial input are well preserved even in the high-noise scenario. In the case of the TST (see figure 4 below), the situation is similar. Again, the AIF profile for the FDCTP data is close to the AIF profile obtained from the CTP data of reduced dimension to a given basis, regardless of the noise level. This is partly due to the fact that the arterial signal with a dynamic change of more than 200 Hounsfield units is the most pronounced of all the TAC curves. In particular, an important aspect

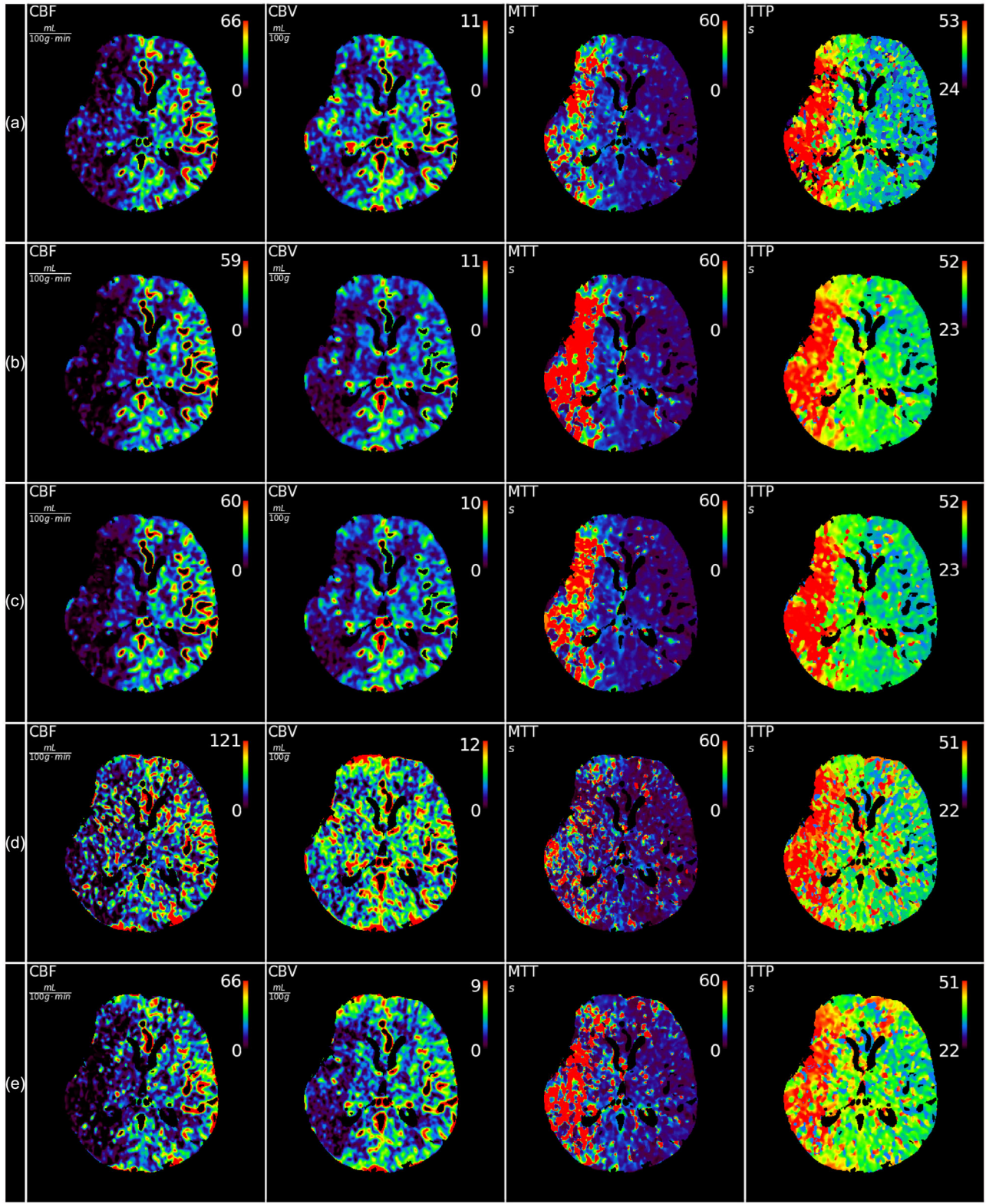

**FIGURE 5** Perfusion maps for patient 1A, $z = 15$. (a) ground truth CTP data; (b) CTP data, dimension reduction to four trigonometric functions; basis $\mathcal{T}_4$; (c) CTP data, dimension reduction to six trigonometric functions, basis $\mathcal{T}_6$; (d) FDCTP data, moderate noise, straightforward method; (e) FDCTP data, moderate noise, TST with the basis of four trigonometric functions $\mathcal{T}_4$

of perfusion imaging is the correct capture of soft tissue dynamics, where the signal can be strongly confounded by the low soft tissue contrast during FDCT acquisition due to the presence of noise. This is apparent when processing FDCTP data by the straightforward method (see figure 5d), where noise overlays important diagnostic information even in a moderate noise scenario.

The TTP perfusion maps calculated from ground truth perfusion data (see Section 2.6) contain more noise than the maps of other perfusion parameters. This is evident, for example, from figure 5a and is related to the fact that TTP is the only parameter for which the stabilized deconvolution operator is not used. This problem was already described in Manhart et al.,[15] where the computation of TTP had to be stabilized using Savitzky–Golay smoothing. In our case, the dimension reduction to the trigonometric basis acts similarly to Savitzky–Golay smoothing to stabilize the TAC shapes. Comparing the perfusion maps in figure 5a with figure 5b, it is clear that the TTP map in figure 5b visually preserves important perfusion patterns but appears smoother. Therefore, the lower TTP correlations in Tables 2–4 when using trigonometric bases are apparently affected by the different levels of smoothness of the compared maps.

When processing the FDCTP data and comparing the TST using the basis $\mathcal{T}_4$ with the straightforward method, we see that the correlations of all perfusion parameters except TTP are on average better for the TST (see Tables 3 and 4). According to this quantitative comparison, the CBF, CBV, and MTT maps show high similarity to the ground truth data for the TST. The TTP correlation coefficients are comparable for both methods. The slightly inferior TTP values may be related to different levels of data smoothing as discussed in the previous section. By visual inspection, it can be concluded that the TST captures all the important information present in the ground truth CT data with less noise than the straightforward method (see, e.g., the comparison of figure 5D with 5E).

At higher noise levels, TST performs much better than the straightforward method (see Table 2). When using the high-noise scenario, the results of the straightforward method deteriorate quickly as they are contaminated by noise, while on perfusion maps that are based on the TST with basis $\mathcal{T}_4$, it is still possible to detect the perfusion deficit, the visualizations are not shown here due to lack of space.

Trigonometric bases and TST proved to be a suitable choice for FDCTP data processing without prior information about the shape of the time attenuation curves. A question, which will be addressed in future work, is how the use of prior knowledge-based information can further improve the presented results. A possible approach is to prepare an orthogonal basis based on the SVD decomposition of a multipatient CTP dataset.[18] The problem is the high variability of attenuation profiles and different contrast agent arrival times. Quite high variability of AIF profiles is visible even for the studied cohort of seven subjects treated in a single medical center, data not shown. As the PCA bases are not shift invariant, different positions of AIF maxima for each individual need to be addressed in prior knowledge-based approaches.

A limitation of the study is the use of simulated FDCTP data instead of actual FDCTP datasets. Using reprojected CTP data with additional noise to create a realistic approximation of FDCTP data was an option to bridge the gap between purely simulated data and patient safety requirements.

From the run times in Section 3.3, it is evident that we have managed to speed up the processing times such that the entire perfusion processing can be performed within 5 min. This was achieved due to the fast parallel GPU implementation of algorithms for CT algebraic reconstruction[28] and optimization of the deconvolution-based perfusion processing.[29] Since the TST with the basis of five functions requires half the reconstructions compared to the straightforward approach, it is faster and less computationally intensive even with significant parallelization of the problem.

The TST with the basis of analytical functions is not necessarily a tool only to study perfusion. Because we use the general basis of trigonometric functions, there is the potential for the method to be used to study general time-resolved CBCT problems. When studying non-periodic motion-related problems, it might be interesting to consider the orthogonal bases of the real polynomials, namely Legendre or Chebyshev polynomials, as a potential basis functions. We would like to work in this direction in the future.

The proposed dimension reduction techniques could be used in dose optimization and noise reduction not only for FDCTP but also for CTP imaging. The reported effective doses for multisweep FDCTP protocols are between 4.6 mSv.[12] and 5.9 mSv.[38] For low-dose CTP protocols, a minimum dose of 2.5 mSv is achievable,[39] while standard protocols for 80 kV work with doses of 3.8 mSv.[40] or 5.0 mSv.[39] Significantly higher doses are reported for higher tube voltages. With the increasing use of CT perfusion scans in practice, data dimension reduction techniques can help to reduce the overall dose to the patient.[41]

## 5 | CONCLUSION

In this study, we have shown that for the analysis of FDCTP data, the TST outperforms the straightforward approach. Using the basis $\mathcal{T}_4$ of four trigonometric functions, noise in the data was noticeably reduced and the resulting perfusion maps were able to accurately capture the brain perfusion status of stroke patients and control subjects. Studies on real data and larger numbers of patients are needed to confirm these results and to deploy FDCT perfusion imaging in clinical practice.

## ACKNOWLEDGMENTS
The work of this paper is partly funded by the Federal Ministry of Education and Research within the Research Campus STIMULATE under grant number 13GW0473A. Vojtěch Kulvait did his part of this work as a post doc at the University of Magdeburg. Since 2022, he is with the Helmholtz-Zentrum Hereon.

Open access funding enabled and organized by Projekt DEAL.

## CONFLICT OF INTEREST
The authors have no conflicts to disclose.

## ORCID
*Vojtěch Kulvait*
https://orcid.org/0000-0003-3279-3757
*Philip Hoelter*
https://orcid.org/0000-0001-9768-9630
*Robert Frysch*
https://orcid.org/0000-0003-2699-9468
*Hana Haseljić*
https://orcid.org/0000-0003-3959-2122
*Georg Rose*  https://orcid.org/0000-0002-2215-150X